\documentclass[a4paper]{jpconf}

\usepackage{citesort}
\usepackage{graphicx}
\usepackage{epstopdf}
\usepackage{amssymb}
\usepackage{lineno}
\usepackage{xspace}
\usepackage{amsmath}
\usepackage{upgreek}
\usepackage{color}

\newcommand{\krl}{\ensuremath{\kern-0.18em}}
\newcommand{\krr}{\ensuremath{\kern-0.09em}}
\newcommand{\tms}{\ensuremath{\kern-0.1em\times\kern-0.2em}}

\newcommand{\ptt}{\ensuremath{p_{\mathrm{T}}}\xspace}

\newcommand{\pb}{Pb--Pb\xspace}
\newcommand{\ppb}{p--Pb\xspace}

\newcommand{\dau}{d--Au\xspace}
\newcommand{\ada}{A--A\xspace}

\newcommand{\rs}[1][7~TeV]{\ensuremath{\sqrt{s}=}~#1\xspace}

\newcommand{\rsnn}[1][2.76~TeV]{\ensuremath{\sqrt{s_{\mathrm{NN}}}=}~#1\xspace}

\newcommand{\gvc}{\ensuremath{\mathrm{GeV}\krl/\krr c}\xspace}

\newcommand{\pion}{\ensuremath{\uppi}\xspace}
\newcommand{\pix}{\ensuremath{\pion^{\pm}}\xspace}

\newcommand{\kx}{\ensuremath{\mathrm{K}^{\pm}}\xspace}
\newcommand{\km}{\ensuremath{\mathrm{K}^{-}}\xspace}
\newcommand{\kp}{\ensuremath{\mathrm{K}^{+}}\xspace}

\newcommand{\ppi}{\ensuremath{\mathrm{p}\kern-0.05em/\krr\pion}\xspace}

\newcommand{\ks}{\ensuremath{\mathrm{K^{*0}}}\xspace}

\newcommand{\ksm}{\ensuremath{\mathrm{K^{*}\krr(892)^{0}}}\xspace}
\newcommand{\ksbm}{\ensuremath{\mathrm{\overline{K}^{*}(892)^{0}}}\xspace}
\newcommand{\ph}{\ensuremath{\upphi}\xspace}
\newcommand{\phm}{\ensuremath{\ph(1020)}\xspace}

\newcommand{\ksk}{\ensuremath{\ks\krl/\mathrm{K}}\xspace}

\newcommand{\kskm}{\ensuremath{\ks\krl/\km}\xspace}

\newcommand{\pks}{\ensuremath{\mathrm{p}\kern-0.1em/\ks}\xspace}
\newcommand{\pksm}{\ensuremath{\mathrm{p}\kern-0.1em/\ksm}\xspace}

\newcommand{\phik}{\ensuremath{\ph\krl/\mathrm{K}}\xspace}

\newcommand{\phikm}{\ensuremath{\ph\krl/\km}\xspace}

\newcommand{\phip}{\ensuremath{\ph\krl/\krr\mathrm{p}}\xspace}

\newcommand{\pphi}{\ensuremath{\mathrm{p}\kern-0.1em/\krl\ph}\xspace}
\newcommand{\pphim}{\ensuremath{\mathrm{p}\kern-0.1em/\krl\phm}\xspace}

\newcommand{\dd}{\ensuremath{\mathrm{d}}}
\newcommand{\mpt}{\ensuremath{\langle\ptt\rangle}\xspace}

\newcommand{\dnc}{\ensuremath{\dd N_{\mathrm{ch}}\kern-0.06em /\kern-0.13em\dd\eta}\xspace}
\newcommand{\dncr}{\ensuremath{(\dnc)^{1/3}}\xspace}

\newcommand{\raa}{\ensuremath{R_{\mathrm{AA}}}\xspace}
\newcommand{\rppb}{\ensuremath{R_{\mathrm{pPb}}}\xspace}

\begin{document}
\title{Resonances as Probes of Heavy-Ion Collisions at ALICE}

\author{A. G. Knospe (for the ALICE Collaboration)}

\address{The University of Texas at Austin, Department of Physics, Austin, TX, USA}

\ead{anders.knospe@cern.ch}


\begin{abstract}
Hadronic resonances serve as unique probes in the study of the hot and dense nuclear matter produced in heavy-ion collisions.  Properties of the hadronic phase of the collision can be extracted from measurements of the suppression of resonance yields.  A comparison of the transverse-momentum spectra of the \phm meson and the proton (which have similar masses) can be used to study particle production mechanisms.  Resonance measurements in pp collisions provide input for tuning QCD-inspired particle production models and serve as reference measurements for other collision systems.  Measurements of resonances in p--Pb collisions allow nuclear effects in the absence of a hot and dense final state to be studied.  The ALICE Collaboration has measured resonances in pp, p--Pb, and Pb--Pb collisions.  These measurements will be discussed and compared to results from other experiments and to theoretical models.
\end{abstract}

Resonances serve as useful probes that allow the characteristics of heavy-ion collisions to be studied at different stages of their evolution.  While the yields of stable hadrons are fixed at chemical freeze-out, the yields of resonances can be modified by hadronic scattering processes after chemical freeze-out~\cite{Bleicher_Stoecker,Markert_thermal,Vogel_Bleicher}.  Regeneration, in which resonance decay products scatter pseudo-elastically through a resonance state (\textit{e.g.}, $\pion\mathrm{K}\rightarrow\ks\rightarrow\pion\mathrm{K}$), can increase the measured yield of the intermediate resonance state without changing the yields of the stable hadrons.  Elastic re-scattering of a resonance decay product can smear the invariant mass resolution and may impede reconstruction of the original resonance.  Pseudo-elastic scattering of a resonance decay product through a different resonance state (\textit{e.g.}, a pion from a \ks decay scattering through a $\uprho$ state) will prevent reconstruction of the first resonance~\cite{Bliecher_Aichelin}.  Regeneration and re-scattering are expected to be most important for $\ptt\lesssim 2$~\gvc~\cite{Bleicher_Stoecker,Bliecher_Aichelin}.  The final resonance yields at kinetic freeze-out will be determined by the chemical freeze-out temperature, the time between chemical and kinetic freeze-out, the resonance lifetime, and the scattering cross sections of its decay products with other hadrons.  Theoretical models that take these effects into account can be used to estimate the properties of the hadronic phase using measured resonance yields (or their ratios to stable particles) as input~\cite{Markert_thermal,Torrieri_thermal,Torrieri_thermal_2001b}.

In addition, partial restoration of chiral symmetry around the phase transition between partonic and hadronic matter may lead to changes in the masses or the widths of resonances~\cite{Brown_Rho,Rapp2009,Brodsky_chiral,Eletsky}.  Mechanisms that determine the shapes of particle \ptt spectra, including the relative strengths of quark recombination~\cite{Fries_Muller_2003,Coalescence_Review_2008} and hydrodynamical effects~\cite{VISH2p1_MCGlb,VISH2p1_MCKLN,KRAKOW}, are studied experimentally using many different particle species; the \phm, a meson with a mass similar to the proton, provides valuable information regarding the effects of mass and baryon number on the shapes of particle \ptt spectra.  Measurements of the nuclear modification factor \raa of hadrons allows the in-medium energy loss of partons to be studied.  Resonance \raa measurements will provide constraints against which the results of theoretical models can be tested.  The effects of ordinary nuclear matter on particle \ptt spectra can be studied through measurement of the nuclear modification factor in \ppb collisions (\rppb).

These proceedings describe recent results from the ALICE experiment related to these topics, with the focus on the \ks (denoting the average of \ksm and \ksbm) and the \ph (abbreviating \phm).  Further details of these results can be found in ~\cite{ALICE_strange_900GeV,ALICE_kstar_phi_7TeV,ALICE_Kstar_phi_PbPb,Bellini_QM2014}.  The main components of the ALICE detector~\cite{ALICE_detector} used in the analyses described here are the Inner Tracking System (for tracking and vertex finding), Time Projection Chamber~\cite{ALICE_TPC} (for tracking and particle identification), the Time-of-Flight detector (for particle identification), and the VZERO detector (used to determine event multiplicity and centrality classes).

The \ks and \ph are measured through invariant-mass reconstruction of their charged hadronic decay channels ($\ks\rightarrow\pix\mathrm{K}^{\mp}$ and $\ph\rightarrow\km\kp$).  The combinatorial background is estimated using an event-mixing technique.  After subtraction of this background, the resonance signals are fitted using a peak fit function (a relativistic Breit-Wigner function for \ks, the convolution of a Breit-Wigner function and a Gaussian for \ph) added to a first- or second-order polynomial, which describes the residual background.  The \ptt-dependent yields, masses, and widths of \ks and \ph are extracted from these fits.  The \ptt spectra are fitted using L\'{e}vy-Tsallis functions for pp and \ppb collisions and Boltzmann-Gibbs blast-wave functions for \pb collisions.  These fit functions are used to extract the mean transverse momentum \mpt and to extrapolate the resonance yields below the lowest measured \ptt bin so that the total \ptt-integrated yield can be calculated.  These quantities have been measured for \pb collisions at \rsnn~\cite{ALICE_Kstar_phi_PbPb}, \ppb collisions at \rsnn[5.02~TeV]~\cite{Bellini_QM2014}, and pp collisions at \rs[0.9, 2.76, and 7~TeV]~\cite{ALICE_strange_900GeV,Bellini_QM2014,ALICE_kstar_phi_7TeV}.

The \ks and \ph masses and widths have been measured as functions of \ptt for central and peripheral \pb collisions at \rsnn~\cite{ALICE_Kstar_phi_PbPb}.  These quantities are consistent with the values extracted from Monte-Carlo simulations (in which particles are produced using HIJING and their interactions with the ALICE detector are simulated using GEANT3) and no centrality dependence is observed.

\begin{figure}[h]
\begin{minipage}{18.5pc}
\includegraphics[width=18.5pc]{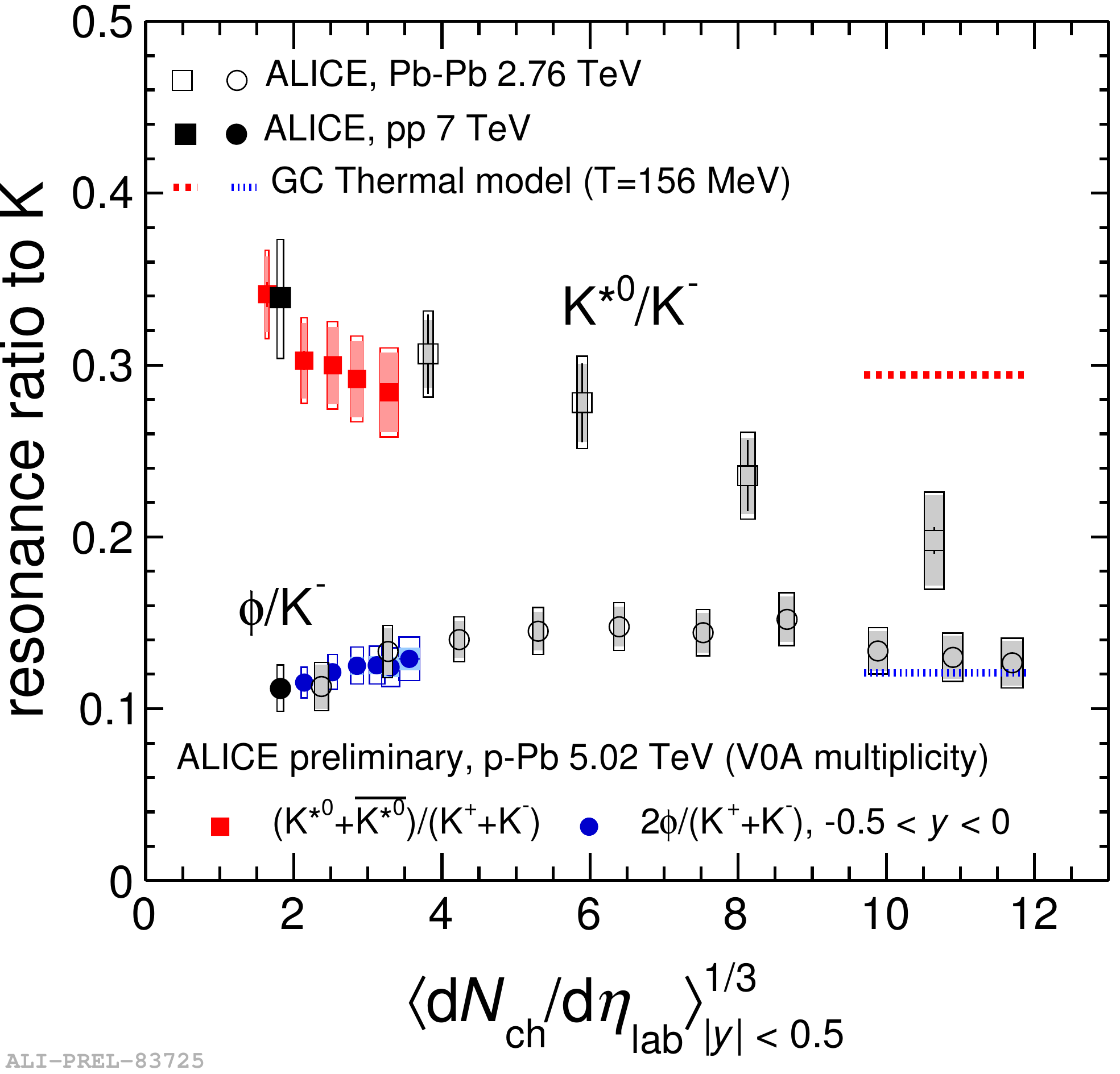}
\end{minipage}
\begin{minipage}{0pc}
\hspace{-3pc}\vspace{14pc}
(a)
\end{minipage}
\begin{minipage}{18.5pc}
\includegraphics[trim=0pc 0pc 4pc 4pc, clip, width=18.5pc]{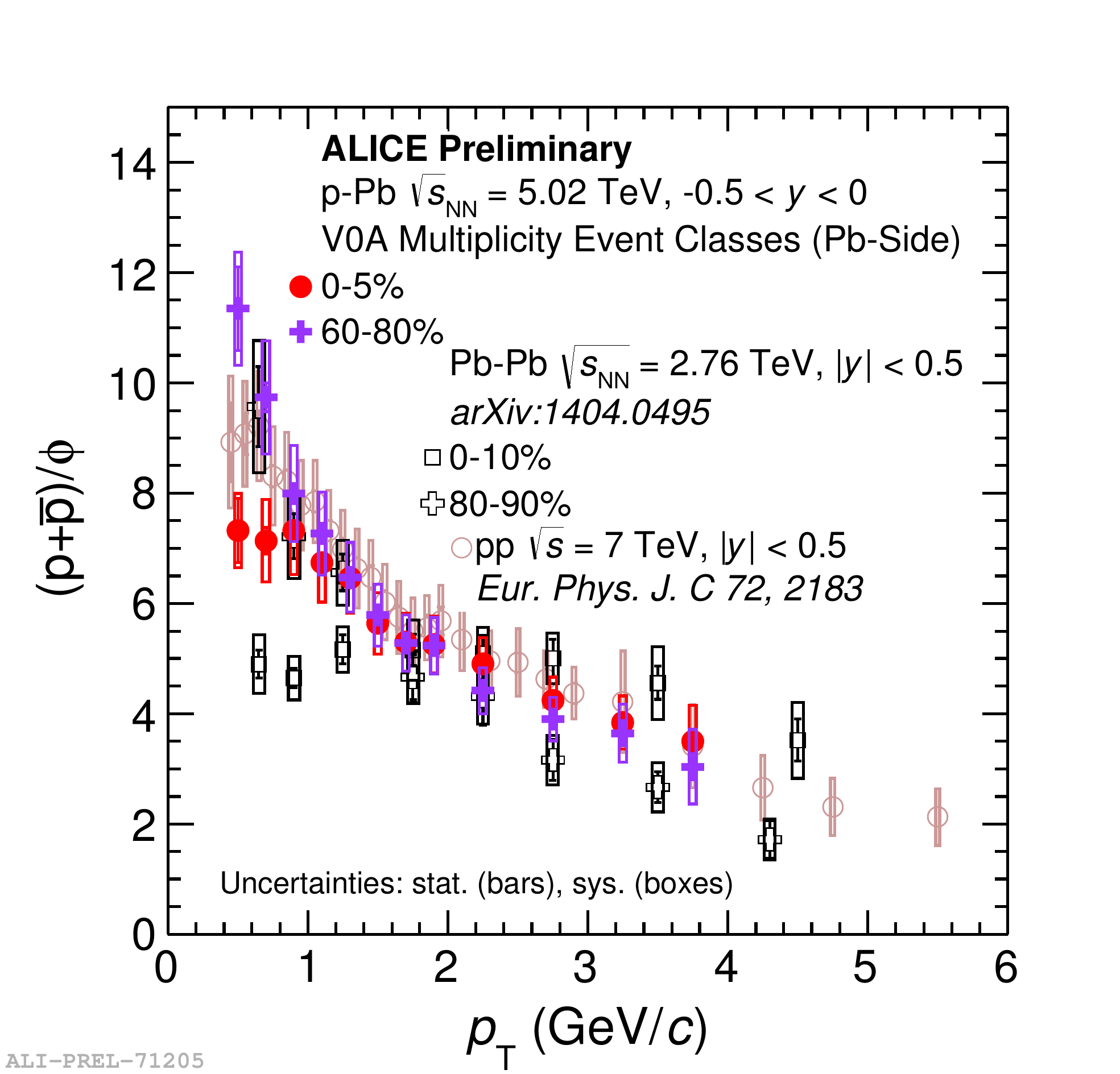}
\end{minipage}
\begin{minipage}{0pc}
\hspace{-15.25pc}\vspace{14pc}
(b)
\end{minipage}
\caption{\label{fig1} For pp, \ppb, and \pb collisions: (a) ratios \ksk and \phik as functions of \dncr, (b): Ratio \pphi as a function of \ptt in different centrality and multiplicity intervals.}
\end{figure}

The ratios of \ptt-integrated particle yields \ksk and \phik are shown in Fig.~\ref{fig1}(a) as a function of \dncr for pp, \ppb, and \pb collisions.  The abscissa \dncr is used as a proxy for the system radius, following a practice used in femtoscopy studies~\cite{Lisa_FemtoscopyReview}.  The \kskm ratio in central collisions is suppressed with respect to the ratio in pp, \ppb, and peripheral \pb collisions.  The \kskm ratio is also suppressed with respect to the value extracted from a thermal model fit of ALICE data~\cite{Stachel_SQM2013} with a chemical freeze-out temperature of 156~MeV and zero baryo-chemical potential (the \ks is excluded from this fit). The centrality-dependent suppression of the \kskm ratio in \pb collisions is consistent with a scenario in which the reconstructible \ks yield is reduced due to re-scattering (with re-scattering being stronger than regeneration).  As described in~\cite{ALICE_Kstar_phi_PbPb}, the measured \kskm ratio in central \pb collisions can be used along with a theoretical model~\cite{Markert_thermal,Torrieri_thermal,Torrieri_thermal_2001b} (a thermal model with resonance yields modified by re-scattering) to obtain an estimate of 2~fm/$c$ for the lower limit of the time between chemical and kinetic freeze-out.  In contrast, the \phik ratio is not suppressed in central collisions and only a weak centrality dependence is observed.  This suggests that re-scattering effects are not important for the \ph, which has a lifetime an order of magnitude longer than the \ks and longer than many estimates of the lifetime of the hadronic phase.  The \ksk and \phik ratios measured in \ppb collisions are consistent with the trends observed in pp and peripheral \pb collisions.  The \kskm ratios measured in pp and \ada collisions at \rsnn[62.4 and 200~GeV]~\cite{STAR_Kstar_2011} follow the same trend as the values observed at LHC energies.  The \phikm ratios observed in pp, \dau, and \ada collisions at \rsnn[200~GeV]~\cite{STAR_phi_2009} are consistent within uncertainties with the values observed at the LHC for similar values of \dncr, although the RHIC measurements tend to be larger than the values observed at LHC.

The \ppi ratio shows a pronounced increase as a function of \ptt for $\ptt<3$~\gvc in \pb collisions~\cite{ALICE_piKp_PbPb}.  Different explanations have been proposed as the cause of this increase.  An increasing baryon-to-meson ratio in this \ptt range suggests quark recombination as the particle production mechanism.  However, hydrodynamical effects could also lead to an increasing \ppi ratio because of the different masses of the two particles.  The \ph, a meson that has a mass similar to the proton (within 9\%), allows the baryon-to-meson ratio to be studied without the added complication of a mass difference.  Fig.~\ref{fig1}(b) shows the \pphi ratio as a function of \ptt for different centrality and multiplicity intervals in pp, \ppb, and \pb collisions~\cite{Bellini_QM2014}.  For central \pb collisions, the ratio is flat for $\ptt<3-4$~\gvc.  This is consistent with hydrodynamical effects: the shapes of the \ptt spectra are determined by the particle masses and not the quark content.  In peripheral \pb collisions, the \pphi ratio takes on an increasingly steep slope, suggesting that the p and \ph may have different production mechanisms in these collisions.  The values and slopes of the \pphi ratio observed in pp and low-multiplicity \ppb collisions are similar to those observed in peripheral \pb collisions.  The \pphi ratio in \ppb collisions does not depend on multiplicity for $\ptt>1$~\gvc.  A possible flattening in this ratio is observed for $\ptt<1$ for high-multiplicity \ppb collisions, which could be interpreted as a hint for the onset of collective behavior in these collisions.

The mean transverse momentum \mpt has been measured for \pix, \kx, (anti)protons, \ks, and \ph in pp, \ppb, and \pb collisions and for $\Lambda$ in \ppb collisions~\cite{ALICE_kstar_phi_7TeV,ALICE_piKp_PbPb,ALICE_Kstar_phi_PbPb,Bellini_QM2014,ALICE_piKpLambda_pPb}.  Mass ordering in \mpt is observed in central \pb collisions.  The \ks, p, and \ph, which have similar masses, are observed to have similar \mpt values.  This is expected from hydrodynamical models.  However, a splitting is observed between the p and \ph for peripheral \pb collisions, with the \ph having large \mpt values.  The same effect is observed in the \pphi ratio.  In \ppb and pp collisions, the mass ordering is only approximate: \mpt for the resonances is larger than \mpt for p and $\Lambda$.  A plot of \mpt as a function of particle mass suggests the possibility of two different trends: one for the mesons (including the resonances) and another for the baryons.  The values of \mpt for \ppb follow different trends from the values for \pb collisions as functions of \dncr~\cite{Bellini_QM2014}.  The values of \mpt for the highest-multiplicity \ppb collisions reach (or even exceed) the values observed for central \pb collisions, despite originating from a smaller system.  This suggests the possibility of different particle production mechanisms in \pb and \ppb collisions.

The nuclear modification factors of these resonances have been computed for \pb collisions at \rsnn[2.76~TeV] (\raa) and \ppb collisions at \rsnn[5.02~TeV] (\rppb, for \ph only)~\cite{Bellini_QM2014}.  For central \pb collisions at high \ptt ($\gtrsim 5$~\gvc), both \ks and \ph are strongly suppressed $(\raa\approx0.15-0.2)$, with similar suppression observed for charged hadrons and identified \pix, \kx, and p.  At low \ptt $(<2$~\gvc), the \ks is more suppressed than charged hadrons in central \pb collisions, which may be an indication that re-scattering effects are of greater importance in this \ptt range.  For $2<\ptt<6$~\gvc in central collisions, the \ph \raa is greater than \raa for the lighter mesons (\pix and \kx), but less than \raa for protons, which have a similar mass.  Since no pp reference \ptt spectra have yet been measured at \rs[5.02~TeV], the reference for \rppb is calculated by interpolating between measured \ph spectra for \rsnn[2.76 and 7~TeV].  The values of \rppb for \ph indicate a moderate Cronin peak (reaching a value of $\approx 1.2$) for $3<\ptt<6$~\gvc.  In contrast, \pix have no significant Cronin peak, while protons appear to have a stronger Cronin peak in the same \ptt range.  For $\ptt\gtrsim 8$~\gvc, \rppb does not deviate from unity for \pix, p, or \ph.

In summary, the \ks and \ph resonance have been measured in different collision systems at LHC energies.  The \ksk ratio is suppressed in central collisions, consistent with re-scattering of its decay daughters, while the \phik ratio is not suppressed due to its longer lifetime.  The \phip ratio is flat for $\ptt<3-4$~\gvc in central \pb collisions, which is expected from hydrodynamical models since the particles have similar masses.  The values of \mpt in \pb and \ppb collisions follow different trends as a function of \dncr and strict mass ordering of \mpt values is violated for \ks, p, \ph, and $\Lambda$ in \ppb collisions.  The nuclear modification factor \raa of resonances at high \ptt shows large suppression, consistent with the suppression observed for stable hadrons.  At high \ptt, the \ph meson \rppb is consistent with unity.

\section*{References}
\bibliography{refs}

\end{document}